\documentclass[aps,pre,showpacs,preprint]{revtex4}
\usepackage{graphicx,amsmath,bm}

\begin{document}

\title{Behavior of $susceptible-infected-susceptible$ epidemics on 
heterogeneous networks with saturation}
\author{Jaewook~Joo} 
\affiliation{Department of Physics, Rutgers University, NJ 08854, USA}
\author{Joel~L.~Lebowitz}
\affiliation{Department of Mathematics and Physics, Rutgers University, 
NJ 08854, USA} 
\date{\today}
\begin{abstract}
We investigate saturation effects in susceptible-infected-susceptible (SIS) models of the spread of epidemics in heterogeneous populations.
The structure of interactions in the population is represented by networks with connectivity distribution $P(k)$, including scale-free (SF) networks with power law distributions $P(k)\sim k^{-\gamma}$.
Considering cases where the transmission of infection between nodes depends on their connectivity, we introduce a saturation function $C(k)$ which reduces the infection transmission rate $\lambda$ across an edge going from a node with high connectivity $k$. A mean field approximation with the neglect of degree-degree correlation then leads to a finite threshold $\lambda_{c}>0$ for SF networks with $2<\gamma \leq 3$. 
We also find, in this approximation, the fraction of infected 
individuals among those with degree $k$ for $\lambda$ close to 
$\lambda_{c}$.
We investigate via computer simulation the contact process on a 
heterogeneous regular lattice and compare the results with those obtained 
from mean field theory with and without neglect of degree-degree correlations.
\pacs{89.75.-k,87.23.Ge,05.70.Ln}
\end{abstract}

\maketitle

\section{Introduction}
Many social, biological, and physical systems can be modeled as networks, i.e., connected graphs with at most 
a single edge between nodes: nodes represent entities and edges represent interaction pathways among the 
nodes~\cite{strogatz:2001,watts:1998}. 
The connectivity pattern in these networks encode information about the structure of the system
~\cite{barabasi:2002,dorogovtsev:2002,newman:2003}. 
An important and much studied feature of these networks is their degree distribution 
$P(k)$, where $P(k)$ is the fraction of nodes of the network that have $k$ 
connections to other nodes.
It was found that many interesting networks such as the internet 
\cite{faloutsos:1999} 
and the patterns of human sexual contacts \cite{stanley:2001} are very heterogeneous 
with approximately "scale-free"(SF) degree distribution: i.e., $P(k) \sim k^{-\gamma}$ 
(power law distribution) with $2< \gamma \leq 3$
~\cite{faloutsos:1999,stanley:2001,barabasi:1999,albert:1999}. 
The study of epidemics in heterogeneous networks is therefore of practical 
importance for the control of the spread of cyber viruses 
and biological epidemics.

The mathematical studies of epidemics on the other hand often make the assumption of 
a homogeneous population~\cite{anderson:1992,diekmann:2000}. 
This means that any infective individual is an equally likely source for the 
further transmission of the disease to other members of the population 
with whom that individual is in contact and vice versa.
The simplest epidemiological model of that kind is the SIS model
~\cite{anderson:1992,diekmann:2000}.
In the SIS model, individuals can only exist in two discrete states; 
healthy (but susceptible) or infected. 
The disease processes are specified as follows: 
Infected individuals become susceptible (healthy) at rate $\delta$, 
independently of their environment. We shall choose time units in which $\delta=1$. 
Susceptible individuals become infected at a rate $\lambda$ multiplied
by the number of infected neighbors, i.e., infected nodes to which they are connected by an edge.
When the web of interactions between individuals is taken to be a regular 
lattice the stochastic process describing this system is the Harris' contact process~\cite{liggett:1985}.

Epidemic behavior in these homogeneous networks, where each node has $z$ neighbors, $P(k)=\delta_{k,z}$, show a "phase transition" as the rate $\lambda$ at which an infected individual infects a susceptible neighbor, is changed; i.e., there exists a critical value $\lambda=\lambda_{c}>0$ below which the only stationary state is a disease-free state (or absorbing phase) and above which there is an endemic infected state (or active phase). 
This can be proven rigorously for the stochastic contact process on a regular lattice, an infinite homogeneous network and is inherited by mean field models based on this process. The mean field critical value, $\lambda^{MF}_{c}$, is proportional to $z^{-1}$, the inverse of the number of ``interacting neighbors''~\cite{liggett:1985,marro:1999}. This mean field $\lambda^{MF}_{c}$ is smaller than that for the contact process. The latter depends not only on the number of neighbors but also on the topology of the lattice (see later) ~\cite{liggett:1985,marro:1999}.

An interesting question then is how to extend these models, which correspond to networks 
with homogeneous connectivity, to real world situations where the number of contacts varies 
greatly from one node to another
~\cite{barabasi:2002,dorogovtsev:2002,newman:2003,faloutsos:1999,stanley:2001,barabasi:1999,albert:1999}.
In such heterogeneous networks, each node has a statistically significant 
probability of having a very large number of 
connections compared to the average connectivity of the network. 
The mean field version of this problem 
was studied in~\cite{anderson:1992,may:2001,pastor:2001a,pastor:2001b,newman:2002,volchenkov:2002} where 
it was shown that the epidemic threshold decreases with increasing 
second moment of the connectivity distribution.
As a result epidemic processes in infinite SF networks with diverging second moment, 
$\gamma \leq 3$, are believed not to 
possess any epidemic threshold below which the infection cannot produce an epidemic 
outbreak or an endemic infected state
\cite{may:2001,pastor:2001a,pastor:2001b,newman:2002,volchenkov:2002}. 

The absence of an epidemic threshold in SF networks makes them very vulnerable.
This remains true even if one takes into account the finite size of 
real systems which of course always have finite second moments. 
In general the epidemic threshold for a heterogeneous network is much smaller than for a homogeneous network with the same average number of contacts~\cite{pastor:2002a}.
The presence of "assortative" or "dissortative" two-point degree correlation in SF networks 
with $2< \gamma \leq 3$
does not appear to alter the absence of epidemic threshold~\cite{pastor:2002b,moreno:2003}.

In these analyses all edges are treated in the same way. There are however 
many situations where there are differences between the ``strength'' 
of different edges. We investigate here cases where the assigned weight 
of an edge between two nodes depends on the connectivity of these nodes. 
We consider in particular saturation effects 
due, for example, to the fact that disease transmission requires a certain   
amount of contact time or "finite time commitment" from both individuals in contact.
This has the effect of lowering the effective connectivity for highly 
connected nodes and thereby decreasing the importance of the ``heavy tails'' 
in $P(k)$. 
Such saturation effects then lead to a finite threshold even 
when the second moment of $P(k)$ diverges. An example in which saturation effects, due to temporal 
limitation of interactions, play an important role is Holling's ``the principle 
of time budget'' in behavioral ecology ~\cite{holling:1959}.

Using mean field approximations appropriate for heterogeneous network 
models~\cite{pastor:2001a,pastor:2001b} of 
epidemics we calculate the critical value $\lambda_{c}$ for different saturation patterns. 
We also find in this mean field approximation, 
which neglects degree-degree correlations between different nodes, 
the behavior of the endemic prevalence $\bar{\rho}_{k}$, i.e., 
the fraction of infected nodes of degree $k$, for $\lambda$ close to $\lambda_{c}$. 
They all have the same behavior for $\lambda \simeq \lambda_{c}$,
$\bar{\rho}_{k}\sim A_{k}(\lambda-\lambda_{c})^{\beta}$,  
with $A_{k}$ increasing with $k$; $\beta=1$ when the third moment is finite. 
This dependence on $k$ is missed by the homogeneous approximation of the contact network.

We then investigate, via numerical simulations, the behavior of the stochastic contact process on a 
regular lattice consisting of nodes with two different degrees. The results are compared with 
mean field approximations with and without neglect of degree-degree correlations.

\section{Mean Field SIS Model with Saturation}

The mean field theory for the contact process, obtained by neglecting correlations between different nodes, 
is described by an equation for 
the density of infected nodes $\rho(t)$, present at time $t$, which can be written as ~\cite{marro:1999}
\begin{equation}
\label{mfeq}
\frac{d\rho(t)}{dt}=-\rho(t)+\lambda z \rho(t)(1-\rho(t))
\end{equation}
where $z$ is the coordination number.
Solving Eq.~(\ref{mfeq}) yields $\rho(t)=(\lambda z-1) \rho(0) e^{(\lambda z -1)t}/
[ \lambda z-1 +\lambda z \rho(0) (e^{(\lambda z-1)t}-1) ]$.
The steady state solution of Eq.(1), obtained as $t \rightarrow \infty$, has 
an epidemic threshold $\lambda^{MF}_{c}=1/z$. 
For $\lambda>\lambda^{MF}_{c}$, any initial infection spreads and becomes 
persistent 
with stationary total prevalence level $\bar{\rho}=(\lambda-\lambda^{MF}_{c})/\lambda$. 
Below the threshold($\lambda<\lambda^{MF}_{c}$), the initial infection dies out 
exponentially fast.
A similar transition occurs for the stochastic contact process 
on a regular lattice with edges between nearest neighbors. 
The critical values $\lambda_{c}$ for the lattice $\bm{Z}^{d}$ are  
$\lambda_{c} \simeq 1.6489$ in one-dimension($d=1$, $z=2$), 
$\lambda_{c} \simeq 0.4122$ in two-dimension($d=2$, $z=4$), etc. 
$z \lambda_{c}$ approaches $z \lambda^{MF}_{c}=1$ as $d \rightarrow \infty$~\cite{marro:1999}.

Consider now a general network with degree distribution $P(k)$. 
Let $\rho_{k}(t)$ be the fraction of the nodes with degree $k$ which are infected at time $t$. We define $\lambda C(k,l)$ as the effective transmission, or infectivity,  rate across an edge going from a node with degree $k$ to a node with degree $l$: $C(k,l)=1$ in the absence of saturation effects.

The mean field equation of the contact process on this network ~\cite{pastor:2001a,pastor:2001b} 
which ignores correlations between the states of the nodes, yields 
the following set of differential equations for $\rho_{k}(t)$
\begin{equation}
\label{sfeq}
\frac{d\rho_{k}(t)}{dt}=-\rho_{k}(t)+\lambda Q_{k}(t)(1-\rho_{k}(t))
\end{equation}
We can interpret $\lambda Q_{k}(t)$ as the effective transmission rate of infection to an uninfected node of degree $k$ by all infected nodes with which it is in contact via any of its $k$ edges, 
\begin{equation}
\label{qeq}
Q_{k}(t)=k\sum_{l}P(l |k)C(k,l)\rho_{l}(t)
\end{equation}
Here $P(l|k)$ is the probability that an edge emerging from a node of degree $k$ has its other end at a node with degree $l$ and $C(k,l)$ is the effective strength of such a bond. We shall now assume further that 
\begin{equation}
\label{nodegree-cor}
 P(l|k)=\frac{lP(l)}{z}
\end{equation}
with $z = \sum_{k}kP(k)$, i.e. random attachment (no degree-degree correlation).

To specify $C(k,l)$ in Eq.~(\ref{qeq}), we make the simplifying assumption that an individual with $k$ contacts spends equal time with each neighbor. 
The effective strength of an edge is then given by a product of ratios of effective connectivity to total connectivity of each node in contact,  
\begin{equation}
\label{edge-weight}
C(k,l)=\frac{C(k)C(l)}{k l}
\end{equation}
with $C(k,l)=1$ corresponding to uniform bond strength.

Eq.~(\ref{sfeq}) can now be written in the form 
\begin{equation}
\label{sfeq-nodegree-cor}
\frac{d \rho_{k}(t)}{dt}=-\rho_{k}(t)
+\lambda(1-\rho_{k}(t))C(k) \Theta(\{\rho(t)\})
\end{equation}
where 
\begin{equation}
\label{theta}
\Theta(\{\rho(t)\})=\frac{\sum_{k}P(k)C(k)\rho_{k}(t)}{z}
\end{equation}
Multiplying Eq.~(\ref{sfeq-nodegree-cor}) by $\frac{C(k)P(k)}{z}$ and summing over $k$ yields
\begin{equation}
\label{theta-eq}
\frac{d \Theta(t)}{dt}=-\Theta(t)+ 
\frac{\lambda \Theta(t)}{z} \sum_{k} P(k) C^{2}(k) ( 1-\rho_{k}(t) ) 
\end{equation}
where $\Theta(t)$ is shorthand for $\Theta(\{\rho(t)\})$.
Given $P(k)$ and $C(k)$, Eqs.~(\ref{sfeq-nodegree-cor}),~(\ref{theta}) and~(\ref{theta-eq}) form a closed set of nonlinear differential equations for the $\rho_{k}(t)$ which can be solved in principle for any given initial values $\{\rho_{j}(0)\}$. They reduce to a single equation, Eq.~(\ref{mfeq}), when $P(k)=\delta_{k,z}$ and $\rho_{z}=\rho$. For a general $P(k)$ the number of variables and equations are infinite.

We are interested primarily in finding stationary solutions $\bar{\rho}_{k}$, 
$0\leq \bar{\rho}_{k} \leq 1$, in which not all $\bar{\rho}_{k}=0$.
(There is of course always one solution corresponding to the uninfected state $\rho_{k}(t)=\rho_{k}(0)=0$ for all $k$.) 
We write the stationary version of Eq.~(\ref{sfeq-nodegree-cor}) in the form 
\begin{equation}
\label{steady-rho}
\bar{\rho}_{k}=\frac{C(k)\lambda \bar{\Theta}}{1+C(k)\lambda \bar{\Theta}}
\end{equation}
Multiplying Eq.~(\ref{steady-rho}) by $\frac{P(k)C(k)}{z}$ and summing over $k$ we get 
\begin{equation}
\label{feq}
\frac{1}{\lambda}=\frac{1}{z}\sum_{k} P(k) 
\Bigl [ \frac{C^{2}(k)}{1+C(k)\lambda \bar{\Theta}} 
\Bigr ] \equiv f(\lambda \bar{\Theta})
\end{equation} 
where $f(x)$ is a monotone decreasing function of $x$. Eq.~(\ref{feq}) will have a (unique) solution $\bar{\Theta}(\lambda)$ different from zero if and only 
if $\lambda>\lambda_{c}$, the epidemic threshold, 
\begin{equation}
\label{threshold}
\lambda_{c}=\frac{1}{f(0)}=\frac{z}{\langle C^{2}(k) \rangle }
\end{equation}
For diverging second moment, $\langle C^{2}(k) \rangle =\infty$, $\lambda_{c}=0$. For a regular lattice with $P(k)=\delta_{k,z}$ and $C(k)=k$, 
we recover the usual mean field result, $\lambda_{c}=\frac{1}{z}$.

Once we have found $\bar{\Theta}(\lambda)>0$ from Eq.~(\ref{feq}) 
we then get $\bar{\rho}_{k}(\lambda)>0$ directly from Eq.~(\ref{steady-rho}) for all $k$ for which $C(k)>0$.
To get the behavior of the $\bar{\rho}_{k}(\lambda)$ for $\lambda \downarrow \lambda_{c}$ we expand the right side of Eq.~(\ref{feq}) in small $\lambda \bar{\Theta}$.
This yields 
\begin{equation}
\label{expansion}
\frac{1}{\lambda} = \frac{1}{\lambda_{c}}
-\frac{\lambda \bar{\Theta}}{z} 
\sum_{k}P(k)C^{3}(k)+ O((\lambda 
\bar{\Theta})^{2})
\end{equation}
Solving for $\bar{\Theta}$ with finite $\langle C^{3}(k) \rangle$, we get 
\begin{equation}
\label{theta-expansion}
\bar{\Theta}(\lambda) = \frac{z}{ \langle C^{3}(k) \rangle } 
( \frac{1}{\lambda \lambda_{c}}-\frac{1}{\lambda^{2}}) + \cdots 
\simeq A(\lambda-\lambda_{c})
\end{equation}
and we obtain from Eq.~(\ref{steady-rho}),
\begin{eqnarray}
\label{rho-expansion}
\bar{\rho}_{k} &\simeq& \lambda_{c} C(k) A (\lambda-\lambda_{c}) \\ 
\bar{\rho} &=& \sum_{k} P(k) \bar{\rho}_{k} \sim \lambda_{c} \langle C(k) \rangle A
(\lambda-\lambda_{c})
\end{eqnarray} 
where $A=\frac{z}{\lambda^{3}_{c} \langle C^{3}(k) \rangle }$.

When $\langle C^{3}(k) \rangle=\infty$ both 
$\bar{\Theta}$ and $\bar{\rho}$ are not differentiable as $\lambda$ approaches $\lambda_{c}$ from above.
For a SF network with $C(k)=k$, $\langle C^{3}(k) \rangle$ is finite only for $\gamma>4$ 
where $\bar{\rho} \sim (\lambda-\lambda_{c})$.
However for SF network with connectivity saturation, 
the range of $\gamma$ where $\langle C^{3}(k) \rangle$ is finite can include all cases 
with $\gamma>2$ so that $z$ is finite. 
The case where $\langle C^{3}(k) \rangle$ is infinite is discussed in the Appendix.

\section{Epidemic threshold in SF network with connectivity saturation} 

To see how connectivity saturation modifies the behavior of epidemics,
 we consider two different types of saturating functions $C(k)$, 
\begin{eqnarray}
C_{I}(k) & = & \, \left \{ 
  \begin{array}{ccr}
    k & \! \text{ if $k<k_{max}$ }\! \\ 
    k_{max} & \! \text{ if $k \geq k_{max}$ }\! 
  \end{array} \right.
\end{eqnarray}
\begin{equation} 
C_{II}(k) = \, \frac{k^{p} k_{max} }{ k_{max}+k^{p}} 
\! \text{ with $ \frac{1}{2}<p \leq 1$ } 
\end{equation} 
where $k$ is total connectivity of a node and $k_{max}$ is a parameter.

\begin{figure}
\begin{center}
\includegraphics[height=10cm,width=10cm]{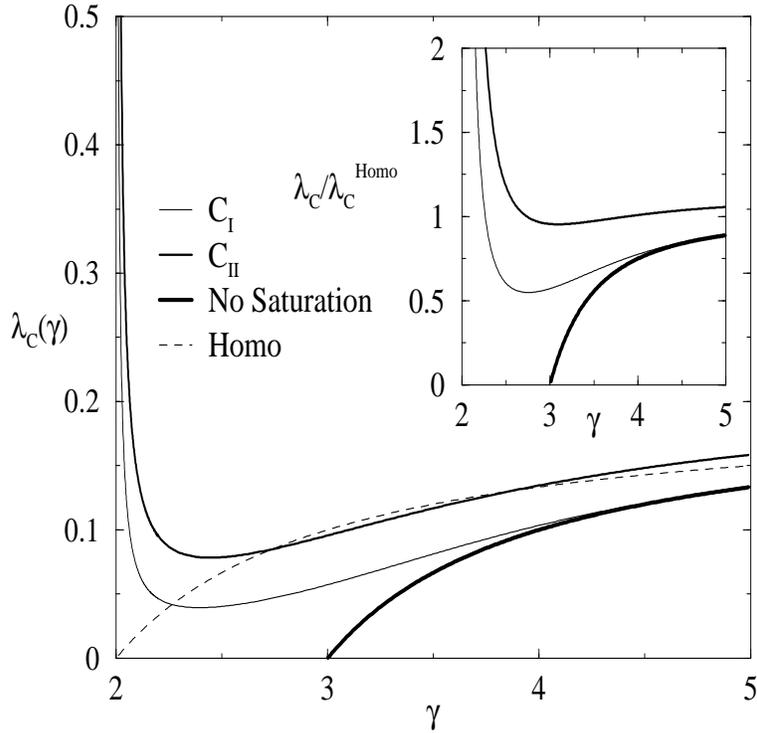}
\caption{\label{fig1}Epidemic thresholds $\lambda_{c}$ as a 
function of $\gamma$. Thin and thick solid lines are drawn for epidemic thresholds from SF 
networks with saturation type $C_{I}$ and $C_{II}$, respectively. 
Fat solid line represents the epidemic threshold from SF network 
without saturation. The dashed line is the epidemic threshold for a homogeneous 
network with coordination number $z=\langle k \rangle$. 
Inset: Ratios of the epidemic thresholds for the SF network with saturation and 
without saturation to that of the homogeneous network.
$k_{max}=100$, $m=5$ and $p=1$ are used.
}
\end{center}
\end{figure}

We replace the sum in Eq.~(\ref{feq}) by an integral over $k$ and carry out calculations of $\lambda_{c}$ for SF networks 
with $P(k)=(\gamma-1)m^{\gamma-1}k^{-\gamma}$ for $k \geq m$, $\gamma>2$. 
After elementary integration we can obtain the epidemic thresholds $\lambda_{c}$ by using Eq.~(\ref{threshold}) and the second order moment $\langle C^{2}(k) \rangle$. 
They are plotted against $\gamma$ in Fig.~\ref{fig1}. 
This figure presents a phase diagram consisting of two phases: 
a disease-free state below each epidemic threshold curve and 
an endemic infection state above each curve.
Note that $\lambda_{c}$ diverges as $\gamma \rightarrow 2$ as can be seen from divergence of $z$ in Eq.(11) when $\langle C^{2}(k) \rangle$ is finite.
This can be understood by noting that as the number of edges increases the effective infection rate for any node decreases.
Note also that $\lambda_{c}=0$ in the absence of saturation for $2<\gamma \leq 3$ because of the divergence of $\langle C^{2}(k) \rangle$.

In the inset of Fig.~\ref{fig1} we compare the epidemic thresholds 
$\lambda_{c}$ of SF networks with saturation with $\lambda^{homo}_{c}$, obtained from a homogeneous network with coordination number $z=\langle k \rangle$.

\begin{figure}
\begin{center}
\includegraphics[height=10cm,width=10cm]{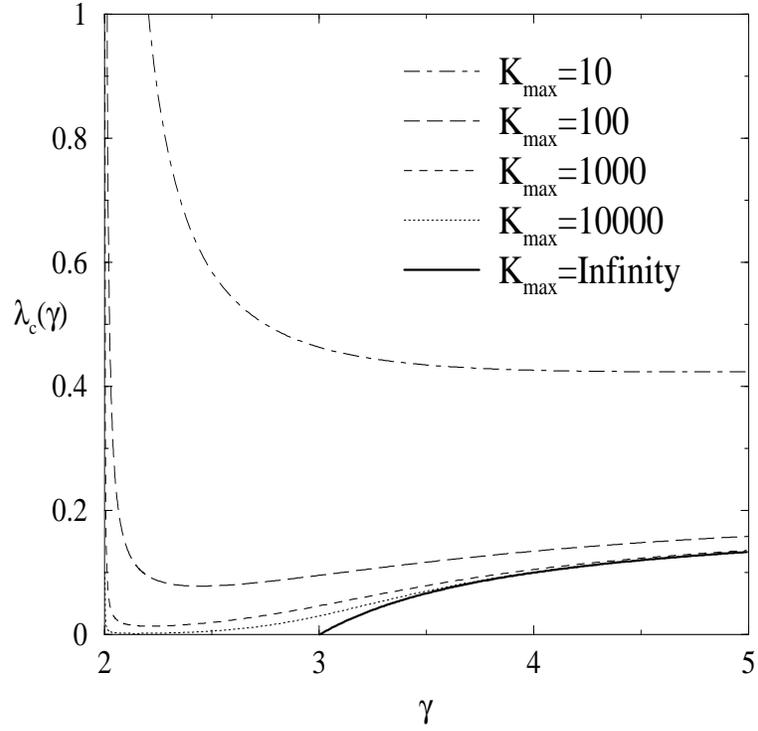}
\caption{\label{fig2}Dependence of epidemic threshold from SF network with saturation 
function $C_{II}$ on the cut-off connectivity $k_{max}$. 
From top to bottom $k_{max}=10$, $10^{2}$, $10^{3}$, $10^{4}$ 
and $\infty$. Here $p=1$ for $C_{II}(k)$.}
\end{center}
\end{figure}

The dependence of the epidemic threshold on $k_{max}$ is plotted in Fig.~\ref{fig2}. 
When $k_{max}$ is finite, the epidemic threshold $\lambda_{c}(\gamma)$ 
is non-zero for $\gamma>2$ and as $k_{max}$ increases $\lambda_{c}(\gamma)$ 
decreases. When $k_{max}=\infty$, the epidemic threshold of SF network without 
saturation is recovered.

\begin{figure}
\begin{center}
\includegraphics[height=10cm,width=10cm]{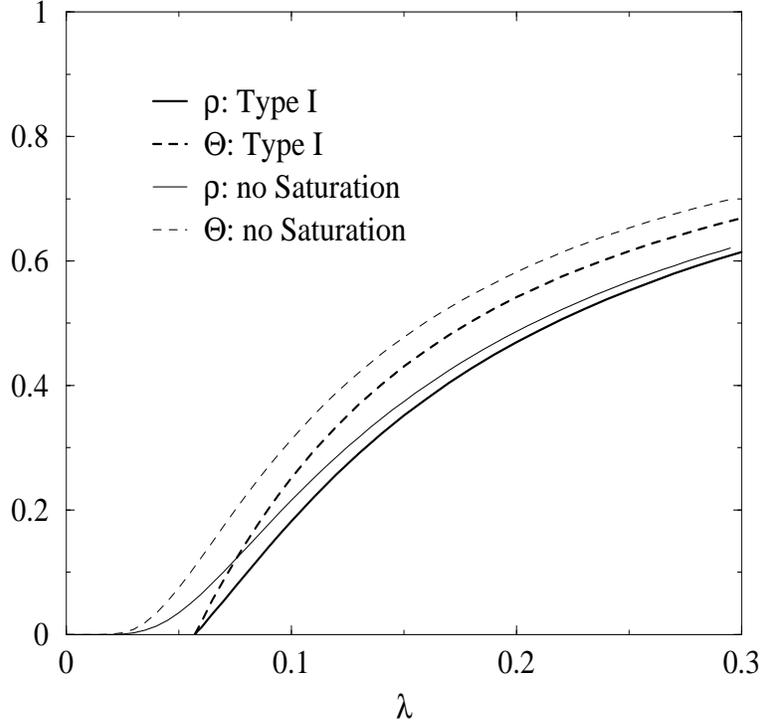}
\caption{\label{fig3}The steady state $\bar{\rho}(\lambda)$ and $\bar{\Theta}(\lambda)$
of the SIS epidemics on the SF network for $\gamma=3$ 
as a function of infectivity $\lambda$. Thick solid(thick dashed) 
lines indicates the steady state $\bar{\rho}^{C_{I}}$ 
($\bar{\Theta}^{C_{I}}$) with Type I saturation. 
Thin lines are for those without saturation. The critical point 
in this particular case is given:  
$\lambda^{C_{I}}_{c}=0.0573$ for Type I saturation and 
$\lambda_{c}=0$ for no saturation.
Here $k_{max}=100$ and $m=5$ are used.
}
\end{center}
\end{figure}

The stationary total prevalence $\bar{\rho}$ for the SF network 
with and without saturation are plotted in Fig.~\ref{fig3}. 
The stationary total prevalence $\bar{\rho}(\lambda)$ with saturation 
is smaller than that without saturation for all $\lambda>0$.
This is because saturation reduces the effective transmission rate of infection to an uninfected node across an edge going from an infected node with high connectivity (see Eq.~(\ref{qeq}) and~(\ref{edge-weight})).

\section{ The contact Process and mean-field approximation in a heterogeneous regular lattice}

\begin{figure}
\begin{center}
\includegraphics[height=10cm,width=10cm]{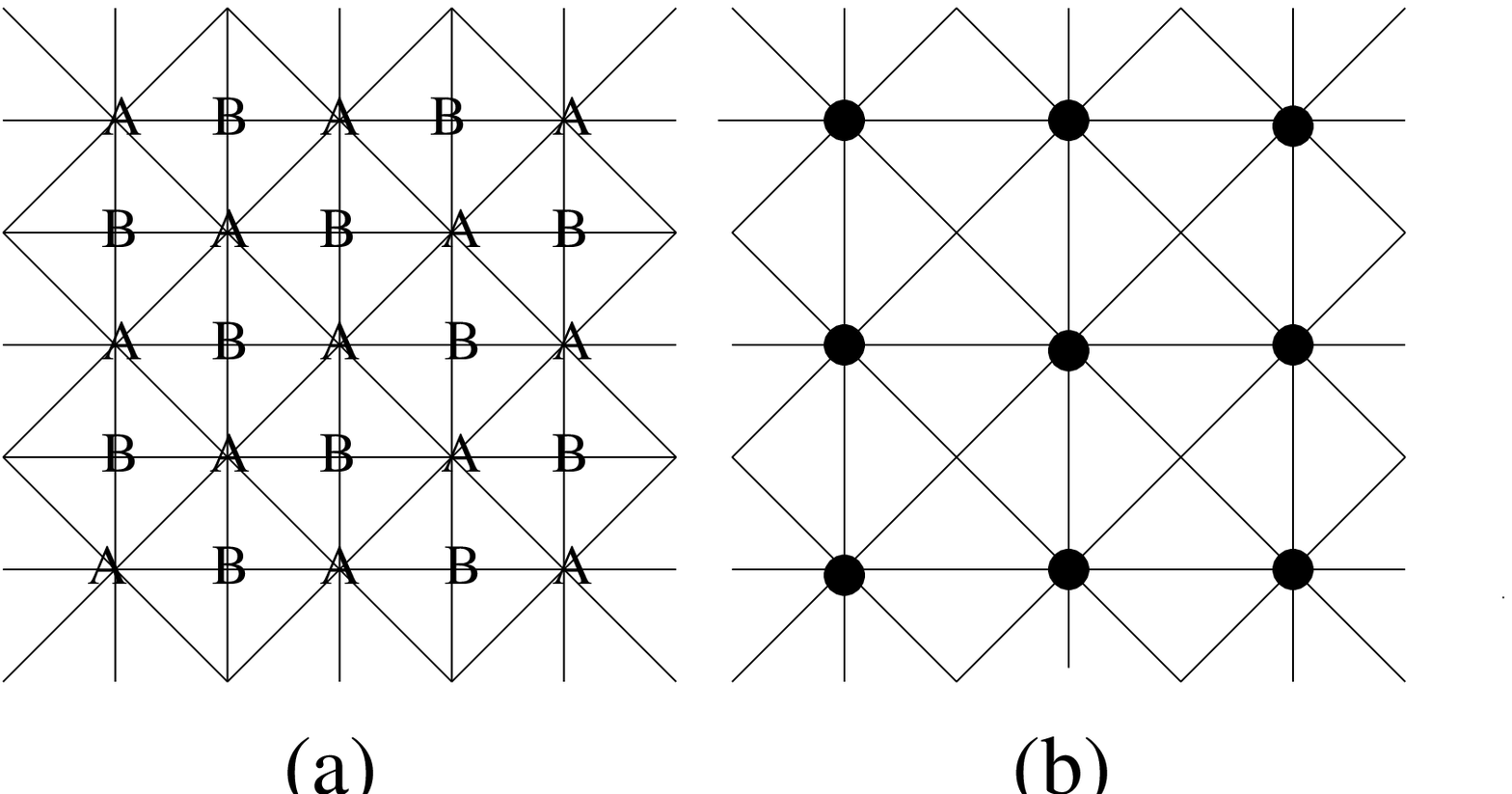}
\caption{\label{fig4}The Topology of lattices. (a) The face-centered 
square (FCS) lattice.
Nodes with type A have 8 degrees while nodes with type B have 4 degrees. 
(b) A homogeneous network with $z=8$.}
\end{center}
\end{figure}

To investigate the effect of heterogeneity and degree-degree correlations 
we investigated the SIS model on the ``face-centered'' square (FCS) lattice with 
two types of nodes. 
Type $A$ nodes, which connect to both nearest and next nearest neighbor
sites, have connectivity $k_{A}=8$ and 
type $B$, which connect only to nearest neighbor sites, have $k_{B}=4$.
For this system $P(8|4)=1$ and $P(4|8)=P(8|8)=1/2$, and 
$P(k_{\gamma}|k_{\alpha})=0$ otherwise: see Fig.~\ref{fig4}.

\begin{figure}
\begin{center}
\includegraphics[height=10cm,width=10cm]{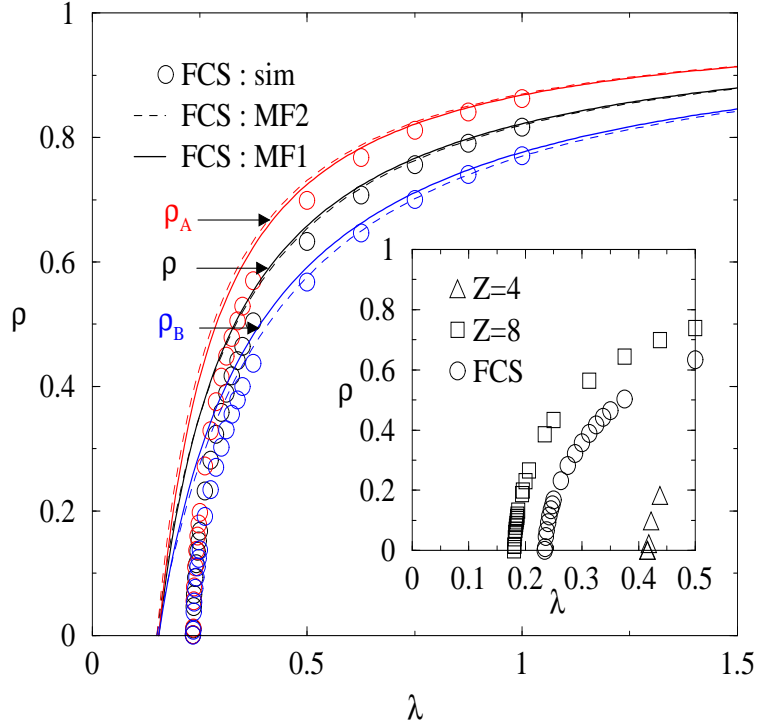}
\caption{\label{fig5}The steady state prevalence $\bar{\rho}$ of $SIS$ in the FCS lattice without saturation as a function of infectivity rate $\lambda$.
$\bar{\rho}_{A}$($\bar{\rho}_{B}$) represents the fraction of infected 
nodes of type A and type B, respectively and $\bar{\rho}=(\bar{\rho}_{A}
+\bar{\rho}_{B})/2$. Simulation (``sim'') data are shown with circles
while mean field results, where ``MF1'' with exact $P(k_{\gamma}|k_{\alpha})$
and ``MF2'' with approximate one using Eq.(4), are given with solid 
and dashed lines, respectively. Inset: Simulation results of the contact process
in the FCS lattice (circles), in the square lattice (triangles) with z=4 and 
in the square lattice with both nearest and next nearest 
neighbor interactions (squares) with z=8. All simulations were performed on 
lattices of size $L^{2}=100^{2}$ with random initial distribution of 
infected nodes.
}
\end{center}
\end{figure}

We carried out computer simulations on this model with no saturation.
The critical point $\lambda_{c}$ and critical exponents, 
$\delta$, $\nu_{\|}$ and $\nu_{\bot}$, were obtained by using the dynamical 
Monte Carlo method~\cite{1979:grassberger}.
As expected there is a critical point $\lambda_{c}=0.23(6)$. 
This value of $\lambda_{c}$ is closer to $\lambda_{c}=0.18(1)$ 
for the homogeneous regular lattice with $z=8$.
than to $\lambda_{c}=0.412$ ~\cite{marro:1999} for the square lattice 
with $z=4$. The homogeneous regular lattice with $z=8$ is 
the square lattice with both nearest neighbor(NN) and next NN bonds (see Fig.~\ref{fig4}). 
The critical exponents appear to be the same as for the square lattice as expected from universality considerations. The phase diagram is plotted in Fig.~\ref{fig5}.

\begin{figure}
\begin{center}
\includegraphics[height=10cm,width=10cm]{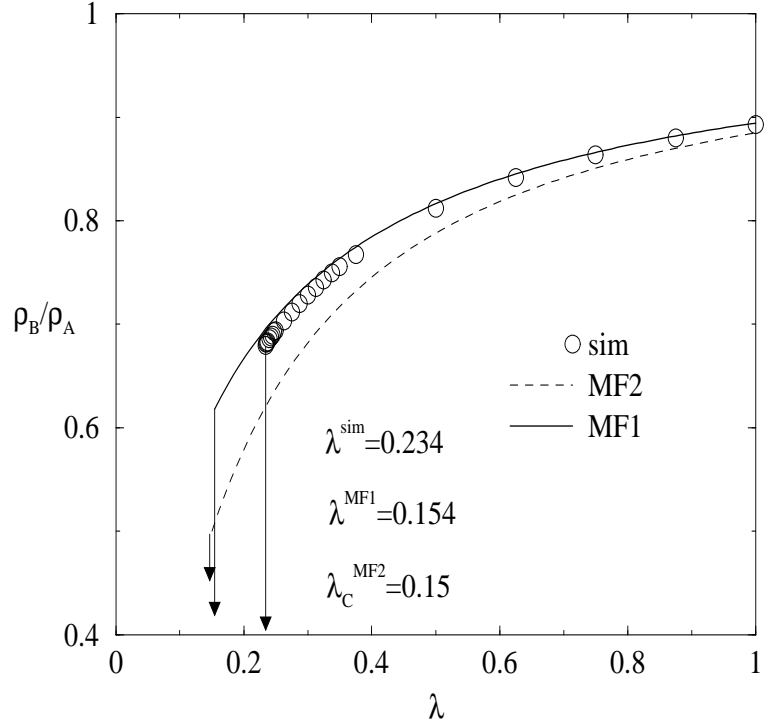}
\caption{\label{fig6}Ratio of $\bar{\rho}_{B}$ to $\bar{\rho}_{A}$ in the FCS lattice 
as a function of infectivity rate $\lambda$. The notation and symbols are the same as in Fig.4.
The arrows point to the critical $\lambda_{c}$ corresponding to each scheme.
}
\label{fig:6}
\end{center}
\end{figure}

We also computed $\bar{\rho}_{A}$ and $\bar{\rho}_{B}$ for $\lambda > \lambda_{c}$. 
The results are plotted in Fig.~\ref{fig6}.  
We can see that nodes with higher connectivity are 
more infected than those with less connectivity at $\lambda$ close to $\lambda_{c}$.

The mean field equations of the SIS on this lattice with the exact 
$P(k_{\gamma}|k_{\alpha})$ are, 
\begin{eqnarray}
\frac{d \rho_{A}}{dt}&=&-\rho_{A}+ 4 \lambda (1-\rho_{A})
( \rho_{A}+\rho_{B} ) \\ 
\frac{d \rho_{B}}{dt}&=&-\rho_{B}+ 4 \lambda \rho_{A}(1-\rho_{B}) 
\end{eqnarray}
The steady state solutions are given, 
\begin{equation}
\bar{\rho}_{A}=\frac{\bar{\rho}_{B}}{4\lambda(1-\bar{\rho}_{B})},
\qquad
\bar{\rho}_{B}= 1-\frac{1}{2\sqrt{\lambda+4\lambda^{2}}}
\end{equation} 
with $\lambda_{c}=(-1+\sqrt{5})/8$. We call this ``MF1''.

The mean field equations with ``no degree-degree correlation'' approximation, 
``MF2'', can be obtained in a similar manner to that given in Eq.(6)-(7). 
This corresponds to putting $P(4|4)=P(4|8)=1/3$ and $P(8|4)=P(8|8)=2/3$.

The results from simulation and the two mean field approximations 
are compared in Fig.~\ref{fig5} and~\ref{fig6} and in Table.~\ref{table1}. Close to the critical point, 
we can approximate the ratio, $\bar{\rho}_{B}/\bar{\rho}_{A}$,
by neglecting nonlinear term in Eq.(19): $\bar{\rho}_{B}/\bar{\rho}_{A} \sim 
4 \lambda_{c} P(8|4)/(1-4 \lambda_{c} P(4|4))$.
In this heterogeneous lattice with only two types of nodes,
``MF2'' with no degree-degree correlation gives the same $\lambda_{c}$ 
but not as good values for $\bar{\rho}_{A}$ and $\bar{\rho}_{B}$
as ``MF1'' with the exact degree-degree correlation.

\section{Concluding Remarks}
In this manuscript we considered the SIS epidemic model on heterogeneous networks with saturation. 
This made the epidemic thresholds finite for SF networks with $2< \gamma \leq 3$. 
We also investigated via computer simulation the stochastic contact process on a heterogeneous 
regular lattice and compared the results with those obtained from mean field theory with and 
without neglect of degree-degree correlations. 
Our considerations extend naturally to other types of heterogeneous networks in which 
the effective strength of an edge depends on the degrees of the nodes which it connects. 
Thus in considering the spread of computer viruses on the 
internet, effects similar to saturation might arise from nodes with high connectivity 
having higher ``firewalls'' around them. 

\begin{acknowledgements}
J.J. thanks DIMACS for support and acknowledges the support of grants NSF DBI 99-82983 and NSF EIA 02-05116. J.J.L. was supported by NSF DMR-01-279-26 and by AFOSR AF 49620-01-1-0154. 
\end{acknowledgements}

\appendix

\section{ Critical behavior of $\bar{\rho}$ when 
$\langle C^{3}(k) \rangle=\infty$ } 

The critical behavior of the steady state $\bar{\Theta}$ and 
$\bar{\rho}$ in the presence of saturation 
can be evaluated by using the third order moment $\langle C^{3}(k) \rangle$. 
When $k_{max}<\infty$, both $\bar{\Theta}$ and $\bar{\rho}$ can be expanded 
in a power series close to the critical point because of $z$, $\langle C^{3}_{I}(k) \rangle$ and  
$\langle C^{3}_{II}(k) \rangle$ being all finite for $\gamma>2$.

When $k_{max} \rightarrow \infty$, $\langle C^{3}(k) \rangle$ may diverge 
and as a result the expression of Eq.(13)-(15) are not valid any more.  
Let us introduce the limiting case of $C_{II}(k)$ when $k_{max}=\infty$,
\begin{equation}
C_{III}(k)=k^{p} \qquad \textrm{for $ \frac{1}{2}< p \leq 1$ } 
\end{equation}
Then, $\langle C^{n}_{III}(k) \rangle = \frac{(\gamma-1)m^{np}}{\gamma-np-1}$ 
for $\gamma > np+1$ while $\langle C^{n}_{III}(k) \rangle=\infty$ for $2<\gamma \leq np+1$.  
When $\gamma>3p+1$, i.e., $\langle C^{3}_{III}(k) \rangle < \infty$, 
$\bar{\Theta}$ and $\bar{\rho}$ at $\lambda$ close to $\lambda_{c}$ 
are equivalent to those in Eq.(13)-(15).

However when $2<\gamma \leq 3p+1$, 
i.e., $\langle C^{3}_{III}(k) \rangle = \infty$, 
$\bar{\rho}$ and $\bar{\Theta}$ are treated in different way and their behaviors 
close to the critical point are presented as follows:
For $C_{III}(k)$ the steady state $\bar{\Theta}^{C_{III}}$ and $\bar{\rho}^{C_{III}}$ 
can be given, 
\begin{eqnarray}
\bar{\Theta}^{C_{III}}&=&\frac{(\gamma-1)m^{p}}{z 
(\gamma-p-1)}
{}_{2}F_{1}[1,\eta-1,\eta,-\frac{1}{\lambda \bar{\Theta} m^{p}}] \\ 
\bar{\rho}^{C_{III}} &=& 
{}_{2}F_{1}[1,\eta,\eta+1,-\frac{1}{\lambda\bar{\Theta}m^{p}}]
\end{eqnarray} 
for non-integer values of $\eta \equiv (\gamma-1)/p$ and for $\gamma>2$. 
After Taylor-expansion in the limit of small $\bar{\Theta}$ 
we obtain (i) $\bar{\rho}^{C_{III}} \simeq \lambda^{(\eta-1)/(2-\eta)}$
for $2<\gamma \leq 2p+1$ (ii) $\bar{\rho}^{C_{III}} \simeq 
\lambda^{1/(\eta-2)}$ for $2p+1<\gamma \leq 3p+1$

\newpage

\begin{table}
\begin{center}
\caption{\label{table1}The critical point $\lambda_{c}$,  
the ratios $\psi_{A(B)}$ and $\phi$, $\psi_{A(B)}=lim_{\lambda \rightarrow \lambda_{c}} 
\frac{\bar{\rho}_{A(B)}(\lambda)}
{\lambda-\lambda_{c}}$ and $\phi=lim_{ \lambda \rightarrow \lambda_{c}} 
\frac{\bar{\rho}_{B}}{\bar{\rho}_{A}}$, of the SIS epidemics 
on the face-centered square lattice.
``Sim'' refers to simulation data without saturation(i.e., $C(k)=k$) while 
``MF1'' and ``MF2'' refers to mean field analysis with exact  
$P(k_{\gamma}|k_{\alpha})$ and approximate one using Eq.(4), respectively.} 
\begin{ruledtabular}
\begin{tabular}{ccccc}
 $$ & $\lambda_{c}$ & $\psi_{A}$ & $\psi_{B}$ & $\phi$ \\
\hline
$Sim$ & 0.23(6) & $\infty$ & $\infty$ & 0.7 \\
$MF1$ & 0.15 & 7.24 & 4.5 & 0.62  \\
$MF2$ & 0.15 & 7.4 & 3.7 & 0.5  \\
\end{tabular}
\end{ruledtabular}
\end{center}
\end{table}

\newpage

\newpage

\newpage

\newpage

\newpage

\newpage

\end{document}